\documentclass[twocolumn,showpacs,preprintnumbers,amsmath,amssymb]{revtex4}

\usepackage{dcolumn}
\usepackage{bm}

\usepackage{afterpage}
\usepackage{epsf}
\usepackage[dvips]{graphics,graphicx,color}
\usepackage{subfigure} 

\usepackage{mathrsfs} 

\usepackage{epsfig}
\usepackage{amsfonts}
\usepackage{amssymb}
\usepackage{amsmath}   
\usepackage{amsthm}
\usepackage{latexsym}

\newtheorem{theorem}{Theorem}[section]

\newtheorem{definition}[theorem]{Definition}

\DeclareMathAlphabet{\mathtensor}{OT1}{cmss}{bx}{n}

\begin{document}

\preprint{APS/123-QED}

\title{Finite Element Formulation of the Bloch Equations with Dipolar
Field Effects}

\author{Louis-S. Bouchard}
 \affiliation{Materials Sciences Division, Lawrence Berkeley National
 Laboratory and Department of Chemistry, University of California,
 Berkeley, CA 94740}

\date{\today}

\begin{abstract}
A Galerkin finite element (FEM) formulation for the Bloch equations with
dipolar field is presented which makes possible the derivation of weak
solutions to the Bloch equations. The FEM formulation has the
advantage that the equations of motion are local in real space,
eliminating the global truncation errors associated with calculations
of the dipolar field in Fourier space. The dipolar field and other
geometric parameters are calculated only once, before the simulation,
and used as an initial condition rather than re-calculated at every
time step of some numerical integration.
\end{abstract}



\maketitle

\newfont{\Bb}{msbm10}

\newcommand{\dotprod}{{\scriptscriptstyle \stackrel{\bullet}{{}}}}

\section{Introduction}

In many modern nuclear magnetic resonance (NMR) and imaging (MRI)
experiments, the Bloch equations with dipolar field are required to
describe the observation of subtle phenomena
~\cite{bib:bouchardenc2003,bib:bouchardjmr,bib:bouchardbone,bib:bouchardxenon,bib:bouchardmrm,bib:warrenscience93,bib:warrenjcp96,bib:warrenscience98,bib:warrenscience96,bib:bowtelljmr92,bib:deville,bib:warrenjcp93,bib:bowtellimaging,bib:bowtellprl96}.
For example, in EDM experiments, dynamic instabilities in highly
polarized liquid xenon have been modeled in this
manner~\cite{bib:bouchardxenon}. Structural imaging at sub-voxel
resolutions in MRI is possible using the long-range magnetic dipole
field~\cite{bib:bouchardjmr,bib:bowtellimaging,bib:bowtellprl96,bib:warrenscience98,bib:warrenscience96}.
In all those cases, considerable insight in understanding the spin
dynamics has been obtained from numerical
simulations~\cite{bib:bouchardxenon,bib:tilman,bib:bouchardjmr}. 

The calculations generally involve the integration of the Bloch equations
with a dipolar field. These are a set of nonlinear first-order partial
differential equations that contain non-local integral operators in
the calculation of the derivative. The dipolar field integral introduces an
$O(N^6)$ operation when calculated at each point in space, making the
calculations very computationally intensive. For this purpose, Enss
and Warren introduced a dramatic speed-up by calculation of the
dipolar field in $k$-space~\cite{bib:tilman}. Any such simplification,
tends to introduce global truncation errors into the numerical
integration, because the dipolar field must be calculated at every
time step using solutions that contain errors accumulated from each
of the previous steps. The cumulation of global truncation may explain
the inability to obtain numerical solutions at long evolution times.

We introduce a finite element (FEM) formulation which may alleviate
these problems, because the dipolar field need not be recalculated at
each time step. This Galerkin formulation of the Bloch equations with
dipolar field was first presented in reference~\cite{bib:bouchardphdthesis}.

\section{Bloch equations with dipolar field}

The secular part of the dipolar field is~\cite{bib:deville,bib:bouchardjmr}:

\begin{equation*}
 \vec{B}(\mathbf{r}) = \int_{\mbox{\Bb R}^3} d^3\mathbf{r}'
 \frac{1-3\cos^2 \theta_{rr'}}{2|\mathbf{r}-\mathbf{r}'|^3} \left[ 3
 M_z(\mathbf{r}') \hat{\mathbf{z}} - \vec{M}(\mathbf{r}') \right] 
\end{equation*}

\noindent Although we use the secular approximation, which is valid in
high magnetic fields, the calculation herein may also be done for the low
field case using the complete dipolar field including all non-secular
terms~\cite{bib:bouchardphdthesis,bib:deville}:

\begin{equation}
 \vec{B}(\mathbf{r}) = \int_{\mbox{\Bb R}^3} d^3\mathbf{r}'
 \frac{1}{|\mathbf{r}-\mathbf{r}'|^3} \left[ \vec{M}(\mathbf{r}') -
 \frac{3 \langle \vec{M}(\mathbf{r}') , \mathbf{r}-\mathbf{r}' \rangle
 (\mathbf{r}-\mathbf{r}')}{|\mathbf{r}-\mathbf{r}'|^2} \right]
\label{eq:exact_ddf}
\end{equation}

\noindent The exact form of the dipolar field is unimportant
here. Chemical shift offsets or radiation damping effects may also be
readily added.

In addition to DDF, diffusion and relaxation, flow effects can also be
accounted for (in this section we write $\vec{M} \equiv
\vec{M}(\mathbf{r},t)$ and  $\vec{B} \equiv \vec{B}_d(\mathbf{r},t)$
for the dipolar field):

\begin{align}
 \frac{\partial \vec{M}}{\partial t} &= \gamma \vec{M} \times \vec{B}
 + \langle \vec{v}(\mathbf{r}) , \nabla \vec{M} \rangle + D \nabla^2
 \vec{M} - \frac{M_x \mathbf{\hat{x}} + M_y
 \mathbf{\hat{y}}}{T_2(\mathbf{r})}  \nonumber \\
 & + \frac{M_0-M_z}{T_1(\mathbf{r})}\mathbf{\hat{z}} \nonumber \\ 
 &= \gamma \vec{M} \times \int d^3\mathbf{r}'
 \frac{1-3\cos^2\theta_{rr'}}{2|\mathbf{r}-\mathbf{r}'|^3} \biggl(
 3M_z(\mathbf{r}',t)\mathbf{\hat{z}}-\vec{M}(\mathbf{r}',t) \biggr)
 \nonumber \\
 & + D \nabla^2 \vec{M} - \frac{M_x\mathbf{\hat{x}} +
 M_y\mathbf{\hat{y}}}{T_2} + \frac{M_0-M_z}{T_1}\mathbf{\hat{z}}
 \nonumber \\
 & + \langle \vec{v} , \nabla \vec{M} \rangle
\end{align}

\noindent This is the set of partial differential equations that is
normally integrated by quadratures to provide numerical
solutions~\cite{bib:tilman,bib:bouchardxenon}.

The diffusion term $\nabla^2 \vec{M}$ requires the existence of the
second derivative. This is a rather strong requirement which is not
usually needed. In the theory of heat conduction, weak solutions to
the heat diffusion equation $\partial T/\partial t = k \nabla^2 T$ often
model physical situations well. In the next section, we investigate the
existence of weak solutions.

\subsection{Weak solutions}

Let $\Omega \subset \mathbb{R}^3$ be the diffusion region and
$\partial \Omega$ be its boundary. The Bloch equations with dipolar
field $\vec{B}(\mathbf{r})$ and magnetic field offset $\delta
\vec{B}(\mathbf{r})$ (to allow for magnetic field inhomogeneities)
are:

\begin{multline}
\frac{\partial \vec{M}}{\partial t} = \gamma \vec{M} \times
(\vec{B}+\delta \vec{B}) - \frac{M_1 \hat{\mathbf{e}}_1 + M_2
  \hat{\mathbf{e}}_2}{T_2} \\
 + \frac{M_0-M_3}{T_1} \hat{\mathbf{e}}_3 + D \nabla^2 \vec{M}
\end{multline}

\noindent are defined for $\mathbf{r} \in \Omega$ and supplemented by
appropriate boundary conditions on $\partial \Omega$, for example
$\hat{\mathbf{n}} \dotprod \nabla \vec{M}=0$ on $\partial \Omega$ in
the case of reflective boundaries. As usual, $\vec{B}$ is the dipolar
field. In component form (component $i$) this reads:

\begin{multline}
\frac{\partial M_i}{\partial t} = \gamma \epsilon_{ijk} M_j (B_k + \delta B_k) - \frac{M_1 \delta_{i1} + M_2 \delta_{i2}}{T_2}  \\
 + \frac{M_0-M_3}{T_1} \delta_{i3} + D \nabla^2 M_i 
\label{eq:bloch_componform}
\end{multline}

Let $V$ be the vector space of test functions on $\Omega$, i.e. all
functions that are continuous, satisfy the boundary conditions on
$\partial \Omega$ and whose first derivative is piecewise continuous
on $\Omega$ (see Quarteroni~\cite{bib:quarteroni} for details). Take a
test function $v_i(\mathbf{r}) \in V$, multiply
Eq.~(\ref{eq:bloch_componform}) by $v_i$ and integrate over $\Omega$:

\begin{multline}
 \int_\Omega  \frac{\partial M_i}{\partial t} v_i d^3\mathbf{r} = \gamma \sum_{j,k} \epsilon_{ijk} \int_\Omega M_j (B_k + \delta B_k) v_i d^3\mathbf{r} \\
  - \frac{1}{T_2} \int_\Omega  \left( \delta_{i1} M_1 v_i + \delta_{i2} M_2 v_i \right) d^3\mathbf{r} \\
 - \frac{\delta_{i3}}{T_1} \int_\Omega  M_3 v_i d^3\mathbf{r} + D \int_\Omega v_i \nabla^2 M_i d^3\mathbf{r} + \frac{M_0}{T_1} \int_\Omega v_i d^3\mathbf{r}
\end{multline}

We proceed to eliminate the second derivative in the diffusion
term. This is done by using the vector identity $\nabla \dotprod (f
\vec{g}) = (\nabla f) \dotprod \vec{g} + f(\nabla \dotprod \vec{g})$
with $f=v_i$ and $\vec{g}=\nabla M_i$. The diffusion term becomes:

\begin{align}
\int_\Omega  v_i \nabla^2 M_i d^3\mathbf{r} =& \int_\Omega \nabla \dotprod (v_i \nabla M_i) d^3\mathbf{r} - \int_\Omega (\nabla v_i) \dotprod (\nabla M_i) d^3\mathbf{r} \nonumber \\
 =& \int_{\partial \Omega} v_i \hat{\mathbf{n}} \dotprod \nabla M_i d^2\mathbf{r} - \int_\Omega (\nabla v_i) \dotprod (\nabla M_i) d^3\mathbf{r} \nonumber \\
 =& - \int_\Omega (\nabla v_i) \dotprod (\nabla M_i) d^3\mathbf{r}
\end{align}

\noindent where in the second line we made use of the divergence
theorem. The surface integral vanishes by virtue of the boundary
conditions $\hat{\mathbf{n}} \dotprod \nabla \vec{M}=0$ on $\partial
\Omega$. Thus, the diffusion term reduces to 

\begin{equation}
 D\int_\Omega  v_i \nabla^2 M_i d^3\mathbf{r} = - D\int_\Omega (\nabla M_i \dotprod \nabla v_i) d^3\mathbf{r}
\end{equation}

\noindent We have derived the following useful result:

\begin{definition}[Weak form of the Bloch equations]
The \textit{weak form} of the Bloch equations---with distant dipolar
field, arbitrary distributions of resonance frequency offsets, and
diffusion within reflective boundaries---is given by (no summation on
i):

\begin{multline}
 \int_\Omega  \frac{\partial M_i}{\partial t} v_i d^3\mathbf{r} =
 \gamma \sum_{j,k} \epsilon_{ijk} \int_\Omega M_j (B_k + \delta B_k)
 v_i d^3\mathbf{r} \\
- \frac{1}{T_2} \int_\Omega  \left( \delta_{i1} M_1 v_i + \delta_{i2}
 M_2 v_i \right) d^3\mathbf{r} \\
 - \frac{\delta_{i3}}{T_1} \int_\Omega  M_3 v_i d^3\mathbf{r} -
 D\int_\Omega (\nabla M_i \dotprod \nabla v_i) d^3\mathbf{r} +
 \frac{M_0}{T_1} \int_\Omega v_i d^3\mathbf{r}
\label{eq:weak_bloch}
\end{multline}

\noindent where $v_i$ is any test function. Any function $M_i$ which satisfies this equation is called a weak solution.
\end{definition}

We note that the degree of differentiability of $M_i$ has been reduced
by half. For the purpose of solving this problem numerically, the
Galerkin method can be introduced.

\begin{definition}[Galerkin formulation] 
Let $V_h$ be a finite dimensional subspace of $V$ of dimension
$N_h$. The Galerkin finite element method consists of finding three
functions $u_i^h(\mathbf{r},t) \in V_h$ ($i=1,2,3$) such that:

\begin{align}
 \int_\Omega \frac{\partial u_i^h}{\partial t} v_i d^3\mathbf{r} = &
 \gamma \sum_{j,k=1}^3 \epsilon_{ijk} \int_\Omega u_j^h (B_k + \delta
 B_k) v_i d^3\mathbf{r} \nonumber  \\
 & - \frac{1}{T_2} \int_\Omega \left( \delta_{i1} u_1^h v_i +
 \delta_{i2} u_2^h v_i \right) d^3\mathbf{r} \nonumber \\
 & - \frac{\delta_{i3}}{T_1} \int_\Omega u_3^h v_i d^3\mathbf{r} - D
 \int_\Omega (\nabla u_i^h \dotprod \nabla v_i) d^3\mathbf{r}
 \nonumber \\
 & + \frac{M_0}{T_1} \int_\Omega v_i d^3\mathbf{r}
\end{align}

\noindent (There is no summation on the index i.) The Galerkin formulation is the matrix form of this problem.

\end{definition}

\noindent \textbf{Remarks.}

\begin{enumerate}
\item The Galerkin method is only an approximation because the
  solution is expanded in terms of a finite-dimensional basis of
  functions.

\item The $u_i^h$ are the best approximations to $M_i$ (in the $L^2$
sense) that the finite-dimensional vector space $V_h$ allows,
i.e. $u_i^h$ is the orthogonal projection of $M_i$ on the subspace
$V_h$. 

\end{enumerate}

Next, we pick a set of basis functions $\left\{ \varphi_i \right\}$
for the vector space $V_h$ and expand the solution in terms of this
basis:\footnote{Since we have lowered the degree of differentiability
 for $M_i$, we have enlarged the set of admissible solutions. For
 example, piecewise linear functions (such as the hat function) do
 not satify Bloch's equations because they are not twice
 differentiable, however, they are perfectly acceptable in
 $V^\circ$.}

\begin{equation}
 u_i^h(\mathbf{r},t) = \sum_{n=1}^{N_h} w_{in}(t) \varphi_n(\mathbf{r})
\end{equation}

\noindent where the functions $w_{in}(t)$ are the unknown coefficients
we must solve for. Taking $v_i = \varphi_m$ yields:

\begin{align}
 & \sum_{n=1}^{N_h} \frac{\partial w_{in}}{\partial t} \int_\Omega
 \varphi_n \varphi_m d^3\mathbf{r} = \nonumber \\
 & = \gamma \sum_{j,k=1}^3 \sum_{n=1}^{N_h} \epsilon_{ijk} w_{jn}
 \int_\Omega \varphi_n \varphi_m (B_k + \delta B_k) d^3\mathbf{r}
 \nonumber \\
 & - \frac{1}{T_2} \sum_{n=1}^{N_h} \left( w_{1n} \delta_{i1}
 \int_\Omega  \varphi_n \varphi_m d^3\mathbf{r}  +  w_{2n} \delta_{i2}
 \int_\Omega  \varphi_n \varphi_m d^3\mathbf{r}  \right) \nonumber \\
 & - \frac{\delta_{i3}}{T_1} \sum_{n=1}^{N_h} w_{3n} \int_\Omega
 \varphi_n \varphi_m d^3\mathbf{r} + D \sum_{n=1}^{N_h} w_{in}
 \int_\Omega \nabla \varphi_n \dotprod \nabla \varphi_m d^3\mathbf{r}
 \nonumber \\
 & +  \frac{M_0}{T_1} \int_\Omega \varphi_m d^3\mathbf{r} 
\end{align}

\noindent In this basis, the dipolar field $B_k$ approximates to:

\begin{multline}
 B_k(\mathbf{r}) = \int_\Omega d^3\mathbf{r}' \frac{1-3\cos^2 \theta_{rr'}}{2|\mathbf{r}-\mathbf{r}'|^3} M_k(\mathbf{r}') a_k  \\
 \approx \sum_{n=1}^{N_h} w_{kn}(t) \int_\Omega d^3\mathbf{r}' \frac{1-3\cos^2 \theta_{rr'}}{2|\mathbf{r}-\mathbf{r}'|^3} a_k \varphi_n(\mathbf{r}')
\end{multline}

\noindent where $a_k = -1$ if $k=1,2$ or $a_k=2$ when $k=3$. Letting
$\| A_{nm} \|$ be the matrix whose elements are $A_{nm} = \int_\Omega
\varphi_n \varphi_m d^3\mathbf{r}$, and assuming it is non-singular,
we denote the inverse of $\| A_{nm} \|$ by $\| C_{nm}\|$. Substitution
of $B_k$ and multiplication of the expression by the matrix $\| C_{nm}
\|$ allows us to solve for the time derivatives:

\begin{widetext}
\begin{align}
 \frac{\partial w_{ip}}{\partial t} = & \gamma \sum_{j,k=1}^3
 \sum_{m,n=1}^{N_h} \epsilon_{ijk} w_{jn} C_{pm}  \int_\Omega
 \varphi_n(\mathbf{r}) \varphi_m(\mathbf{r}) \delta B_k(\mathbf{r})
 d^3\mathbf{r} + \gamma \sum_{j,k=1}^3 \sum_{m,n,q=1}^{N_h}
 \epsilon_{ijk} w_{jn} w_{kq} C_{pm}  \int_\Omega
 \varphi_n(\mathbf{r}) \varphi_m(\mathbf{r})  \nonumber \\
 & \times  \left[ \int_\Omega  \varphi_q(\mathbf{r}') \frac{1-3\cos^2\theta_{rr'}}{2|\mathbf{r}-\mathbf{r}'|^3} a_k d^3\mathbf{r}' \right] d^3\mathbf{r} - \frac{1}{T_2} \sum_{m=1}^{N_h} C_{pm}  \left( \delta_{i1} \sum_{n=1}^{N_h} w_{1n} A_{nm} + \delta_{i2} \sum_{n=1}^{N_h} w_{2n} A_{nm} \right) \nonumber \\
 & - \frac{\delta_{i3}}{T_1} \sum_{m,n=1}^{N_h} w_{3n} C_{pm} A_{nm} + D \sum_{m,n=1}^{N_h} w_{in} C_{pm} \int_\Omega \nabla \varphi_n \dotprod \nabla \varphi_m d^3\mathbf{r} +  \frac{M_0}{T_1} \sum_{m=1}^{N_h} C_{pm} \int_\Omega \varphi_m(\mathbf{r}) d^3\mathbf{r} 
\end{align}
\end{widetext}

\noindent Then, with the following abbreviations,

\begin{align}
 H_m =&  \int_\Omega \varphi_m(\mathbf{r}) d^3\mathbf{r} , \qquad
 G_{nm} = \int_\Omega \nabla \varphi_n(\mathbf{r}) \dotprod \nabla
 \varphi_m(\mathbf{r}) d^3\mathbf{r} \nonumber \\
 R_{knm} =&  \int_\Omega \delta B_k(\mathbf{r}) \varphi_n(\mathbf{r})
 \varphi_m(\mathbf{r}) d^3\mathbf{r}  \nonumber \\
T_{knmq} =& a_k \int_\Omega  \varphi_n(\mathbf{r})
 \varphi_m(\mathbf{r}) \left[ \int_\Omega \varphi_q(\mathbf{r}')
 \frac{1-3\cos^2\theta_{rr'}}{2|\mathbf{r}-\mathbf{r}'|^3}
 d^3\mathbf{r}' \right] d^3\mathbf{r}
\end{align}

\noindent the expression takes a particularly simple form,

\begin{align}
 \frac{\partial w_{ip}}{\partial t}  =& \gamma \sum_{j,k=1}^3
 \sum_{m,n=1}^{N_h} \epsilon_{ijk} w_{jn} C_{pm} (\sum_{q=1}^{N_h}
 w_{kq} T_{knmq}+R_{knm}) \nonumber \\
 & - \frac{1}{T_2} \left( \delta_{i1} w_{1p} + \delta_{i2} w_{2p}
 \right) \nonumber \\
 & - \frac{1}{T_1} \delta_{i3} w_{3p} + D \sum_{m,n=1}^{N_h} w_{in}
 C_{pm} G_{nm}  \nonumber \\
 &+  \frac{1}{T_1} M_0 \sum_{m=1}^{N_h} C_{pm} H_m .
\end{align}

This is a set of coupled ODEs, involving linear and bilinear forms in
the unknowns $w_{ij}$, with constant coefficients $R_{knm}$, $C_{pm}$,
$A_{nm}$, $H_m$, $G_{nm}$ and $T_{knmq}$. These constants can be
calculated \textit{a priori} using the choice of basis functions
$\left\{ \varphi_i \right\}$. One may then solve for the
time-dependent coefficients $w_{ip}$ (there are $3 \times N_h$ of
them) using conventional ODE methods (see, for example,
Quarteroni~\cite{bib:quarteroni}).

\section{Conclusion}

We have shown how to derive the Galerkin formulation of the Bloch
equation. The main advantage to this approach is in calculating
time-evolution of magnetic moments in the presence of a strong dipolar
field, in which case, we believe that error propagation from repeated
dipolar field calculations should be
mitigated~\cite{bib:bouchardphdthesis,bib:tilman,bib:bouchardxenon}.
In contrast with the direct integration of the Bloch equations using
the Fourier method~\cite{bib:tilman,bib:bouchardxenon}, there is no
need to calculate a dipolar field at every point in time; hence, there
is no propagation of global truncation errors associated with the
Fourier approximation of the dipolar field. This set of equations,
while still non-linear, does not involve a non-local integral operator in
the calculation of the derivative at each time step.


\bibliographystyle{unsrt}
\bibliography{dbase8}

\end{document}